# Characterizing the highly cited articles: a large-scale bibliometric analysis of the top 1% most cited research

Pablo Dorta-González [a,*], Yolanda Santana-Jiménez [b]

[a] Universidad de Las Palmas de Gran Canaria, TiDES Research Institute, Campus de Tafira, 35017 Las Palmas de Gran Canaria, Spain. *E-mail*: pablo.dorta@ulpgc.es

[b] Universidad de Las Palmas de Gran Canaria, Departamento de Métodos Cuantitativos en Economía y Gestión, Campus de Tafira, 35017 Las Palmas de Gran Canaria, Spain. *E-mail*: yolanda.santana@ulpgc.es

[*] Corresponding author

**Abstract**

We conducted a large-scale analysis of around 10,000 scientific articles, from the period 2007-2016, to study the bibliometric or formal aspects influencing citations. A transversal analysis was conducted disaggregating the articles into more than one hundred scientific areas and two groups, one experimental and one control, each with a random sample of around five thousand documents. The experimental group comprised a random sample of the top 1% most cited articles in each field and year of publication (highly cited articles), and the control group a random sample of the remaining articles in the Journal Citation Reports (science and social science citation indexes in the Web of Science database).

As the main result, highly cited articles differ from non-highly cited articles in most of the bibliometric aspects considered. There are significant differences, below the 0.01 level, between the groups of articles in many variables and areas. The highly cited articles are published in journals of higher impact factor (33 percentile points above) and have 25% higher co-authorship. The highly cited articles are also longer in terms of number of pages (10% higher) and bibliographical references (35% more). Finally, highly cited articles have slightly shorter titles (3% lower) but, contrastingly, longer abstracts (10% higher).

**Introduction**

It is well known that about 20% of papers obtain more than 80% of citations, while other papers are either not cited at all or are infrequently cited (Garfield 2006). Based on this, when a particular paper is cited more frequently than others of a similar topic and age, it is usually concluded that it has a higher quality compared to the other papers (Bornmann et al. 2012). However, there are other reasons why researchers cite papers: to support their own claims, methodology or findings (supportive citations); to present different points of view; and even to criticize the cited paper (Aksnes 2003; Harwood 2008).

Among the factors influencing the number of citations, Tahamtan, Askar Safipour and Ahamdzadeh (2016) identified three general dimensions: *(i) Paper related aspects*: quality, novelty, interest, field and topic, typology, study design, methodology, results and discussion, figures and appendices, titles and abstracts, references, length, age, early citation, and accessibility; *(ii) Journal related aspects*: journal impact factor, language, scope, and visibility; and *(iii) Author(s) related aspects*: number of authors, reputation, academic category, self-citations, international and national collaboration, country, gender and age, productivity, and funding.

Some authors have studied the *scientific aspects* influencing citations (Buela-Casal and Zych 2010; Callaham et al. 2002; Patterson and Harris 2009; Stremersch et al. 2007). In addition to the quality of the paper, the methodology (Bhandari et al. 2007) together with the novelty of the subject and the popularity of the topic (Chen 2012; Peng and Zhu 2012) seem to be the main scientific aspects influencing citations.

However, in this paper we study the *bibliometric or formal aspects* influencing citations. We restrict the analysis to research articles in order to avoid the typology bias. It is well known that some types of documents receive more citation than others. Generally, review articles receive more citations than research articles (Biscaro and Giupponi 2014; Fu and Aliferis 2010).

Our study differs from others in the literature in various ways. We conducted a large-scale transversal analysis of around 10,000 papers disaggregated into more than one hundred scientific areas. We used a novel methodology comparing two different groups of papers, those highly cited and those not highly cited. The population analysed are the research articles published in the period 2007-2016 in journals of the Journal Citation Reports (science and social science citation indexes in the Web of Science database). Our aim was to verify the existence or otherwise of bibliometric aspects which displayed significant differences between the group of top 1% most cited articles according to their field and year of publication (highly cited articles) and the remaining articles. In each of these two groups of articles, we considered a random sample of around five thousand documents.

**Bibliometric or formal aspects influencing citations: state of the art**

There is no strong evidence in the literature in favour of the thesis that some formal aspects contribute to a paper achieving more citations. The results mainly depend on the methodology employed and there is no consensus with respect to some of these aspects about the real effect on citations. A brief revision of the most relevant bibliometric factors considered in the literature is described below.

*Field and age*

The number of citations varies according to the characteristics of the disciplines and topics (Costas et al. 2009; Dorta-González et al. 2014). Hot topics usually attract more attention and receive more citations (Fu and Aliferis 2010), but the size of the literature (number of papers published in the field) also contributes to the number of citations a paper receives (Biscaro and Giupponi 2014).

With respect to the age of the article, in general the number of citations increases in the first years after publication before reaching a peak and then gradually decreasing over time (Dorta-González and Dorta-González 2013b). One possible reason is that the information becomes increasingly outdated and obsolete (Barnett and Fink 2008).

*Co-authorship and visibility*

The number of authors indicates the extent of the scientific collaboration. Papers with more authors are more likely to obtain a higher number of self-citations, external citations and visibility (Biscaro and Giupponi 2014; Peng and Zhu 2012).

To increase this visibility, researchers also try to publish their papers in high impact journals (to reach more readers and become more frequently cited). The *journal impact factor* (JIF) can be considered a proxy of visibility and prestige, which are of high importance for a document to be cited (Dorta-González et al. 2017; Dorta-González and Santana-Jiménez 2018). Publishing papers in high impact journals would result in more citations than publishing in low impact ones (Aksnes 2003; Callaham et al. 2002; Fu and Aliferis 2010; Garner et al. 2014; Peng and Zhu 2012). However, the impact factor is a consequence of citations, and is often considered a cause of citations. Thus, considering this aspect in predicting citations is controversial (Bhandari et al. 2007).

*Length of the document and references*

The length of a paper is among the factors increasing the number of citations (Falagas et al. 2013; Peng and Zhu 2012; van Wesel et al. 2014). This might be due to the fact that longer papers contain more information. However, some other studies show there is no relationship between the length of a paper and the number of citations (Royle et al. 2013; Walters 2006).

The list of bibliographical references in a paper can be considered the knowledge of the author(s) about the literature. Thus, the number of references, their prestige as well as the variety of the references in a paper increase the frequency of citation (Biscaro and Giupponi 2014; Chen 2012; Falagas et al. 2013; van Wesel et al. 2014).

*Title and abstract*

The characteristics of the title and abstract are not identified as determinant for citations in the literature. The title affects the number of downloads more than the number of citations. Papers with titles in question form are downloaded more than those with descriptive titles, but they are less frequently cited (Jamali and Nikzad 2011). Furthermore, titles with two components separated by a colon increases the number of citations (Jacques and Sebire 2010). Other punctuation marks such as a comma also increase citations (Buter and van Raan 2011).

Although some authors did not find a significant correlation between title length and citations (Jamali and Nikzad 2011; Rostami et al. 2014), others maintain that the title length negatively affects citations, in that longer titles receive less citations than shorter one (Stremersch et al. 2015; Subotic and Mukherjee 2014).

Finally, papers with longer abstracts receive more citations (van Wesel et al. 2014).

**Methodology**

The on-line search application of the *Web of Science* database, currently managed by Clarivate Analytics and available at apps.webofknowledge.com, was used for the data search.

Two citation indexes in the Journal Citation Reports were selected (Science Citation Index Expanded and Social Sciences Citation Index). In addition, the basic search option was employed with the following search criteria: Document Types = (Article) AND Year Published = (2007-2016). This basic search resulted in a total population of around ten millions research articles published between 2007 and 2016.

Regarding the design of the research, two groups of research articles were considered, an experimental and a control group. The experimental group was filtered using the Essential Science Indicators (ESI) Top Papers criterion, refining to the highly cited papers category which uses the ESI to locate the top 1% most cited documents according to their field and year of publication. This search resulted in a total of 99,479 *Highly Cited Articles* (HCA).

Then, 5,000 of these HCA articles were randomly selected. After discarding some anomalous documents with empty relevant data, a total of 4,956 articles remained. This sample represents 4.98% of the total population in the HCA group.

For the control group, the HCA were first excluded and then a simple random sample of 5,000 articles was made, resulting in 4,998 *Non-Highly Cited Articles* (NHCA) after discarding an anomalous pair of data.

Therefore, the total sample size was n = 9,954 research articles published between 2007 and 2016. For this random sample, the following variables were exported directly from the database: Author, Year, Title, Abstract, Source, Page Count, Times Cited, Cited Reference Count, and Research Areas.

In the disaggregated analysis, of the 137 research areas in the Journal Citation Reports, those in which the number of articles in either of the two groups (HCA and NHCA) was less than five cases were discarded. This is because we consider there are not enough data to draw conclusions. After this discarding process, the final number of research areas in the disaggregated analysis was 107.

We also download the JIF from the Journal Citation Reports. Web of Science uses a journal classification system where each journal is assigned to one or several subject categories. According to the JIF, each journal is placed in a percentile within each category. In this paper, we used the best percentile for each journal, that is, the highest of them all. This is the reason why the median is above the 50th percentile, even in the NHCA group.

Finally, we linked the two datasets by the journal. Both the search for the data and its export to the dataset were done during the first week of September 2017.

As the main statistical tools, we used the median and a non-parametric median test. The median is the value separating the higher half from the lower half of a data sample. That is, the middle value of a data set. The basic advantage of the median in describing data compared to the mean is that it is not skewed so much by extremely large or small values, and so it may give a better idea of a typical value. Because of this, the median is of central importance in robust statistics.

Finally, a non-parametric median test was chosen to compare the HCA and NHCA groups since the variables considered in the study do not follow a normal distribution. The non-parametric median test is a statistical tool that tests the null hypothesis that the medians of the populations from which two or more samples are drawn are identical.

**Results and discussion**

*Data distribution and linear correlation between variables*

The hypothesis of normality was rejected for all the analysed variables. Normal contrasts were performed and the frequency histograms corroborated asymmetry and distributions very far from the normal. For this reason, it was decided to use the median in this paper as a measure of central tendency, which is quite common in bibliometric studies (Dorta-González and Dorta-González 2013a).

In relation to independence between the variables analysed, within the 107 areas linear correlations higher than 0.5 were found only between number of references and number of pages, and usually the HCA group had a higher coefficient. Pearson's correlation coefficients for the 10 research areas with the largest sample sizes are shown in Table 1. As can be observed, in the HCA group these coefficients are above 0.58 in most of the areas (7 out of 10), but in only two 2 areas in the NHCA group. Therefore, in the highly cited group, the longest papers are supported by a greater number of references.

[Table 1 about here]

However, no linear correlations were found between any other pair of variables analysed. Interestingly, although the number of authors might be expected to have an impact on both article length and number of citations, no correlation was found in this study between number of authors and number of pages, nor between number of authors and JIF.

*Medians by groups of papers, and equality of median tests in aggregated areas*

A non-parametric median test was chosen to compare the HCA and NHCA groups as the variables in this study do not follow a normal distribution (Table 2). For all the variables analysed there are significant differences, below the 0.01 level, between the groups of articles (HCA and NHCA). The highly cited articles are published in higher impact factor journals (33 percentile points higher), have more authors (25% higher) and are longer in terms of number of pages (10% higher) and bibliographical references (35% higher). In addition, highly cited articles have slightly shorter titles (3% lower) but, contrastingly, longer abstracts (10% higher).

[Table 2 about here]

*Medians by groups of papers, and equality-of-medians tests in disaggregated areas*

In order to reduce the field effect, Table 3 analyses the previous aspects but disaggregating for each of the 107 research areas. This information is also summarized in Figure 1. Note that the medians are higher for the JIF percentile in the HCA group in all research areas. Moreover, in the HCA group the medians are clearly higher in most of the research areas for the rest of the aspects except for length of title. That is, the highly cited articles are, in general, more extensive in number of pages, which is the result of the work of a greater number of authors who reference a greater number of documents. In addition, highly cited articles have slightly shorter titles but, contrastingly, longer abstracts.

[Table 3 and Figure 1 about here]

A summary of Table 3 in relation to the significance level is presented in Table 4. Note that, in general, the percentage of research areas with significant differences between the two groups of papers increases when only considering the 30 research areas (of the total of 107) which had more than 50 papers in both the HCA and NHCA.

[Table 4 about here]

From highest to lowest significance in the results, the following comments can be made about Table 4.

*Journal Impact Factor percentile*

Of the 107 areas considered, 79 display differences between the median percentiles corresponding to the HCA and NHCA groups at a significance level of 1%. If a significance level of 5% is considered, the number of areas showing differences between medians rises to 86. When large samples are considered, the equality-of-medians hypothesis is rejected in all research areas with more than 50 observations in each group. In all cases, the percentile is higher in the HCA group. This result is strong evidence in favour of the hypothesis that publishing in journals with a high impact factor contributes to achieving more citations for a paper.

*Number of authors*

In 36 of 107 areas, there exist differences between the median number of authors corresponding to the HCA and NHCA groups at a significance level of 1%. If a significance level of 5% is considered, the number of areas showing differences between rises to 51. When large samples are considered, the equality-of-medians hypothesis is rejected in 80% of the research areas with more than 50 observations in each group. In all cases, the median number of authors is higher in the HCA group. This result is strong evidence in favour of the hypothesis that collaborations contribute to a paper achieving more citations.

*Number of characters in abstract and titles*

Most people find publications nowadays via Google Scholar or other online sources. The search algorithms used by Google and other search engines assign more importance to words appearing in a title compared with an abstract or the body text of a paper. If the article title includes keywords that other researchers in the field are likely to search for, then the document is much more likely to show up on the search returns. From a bibliometric perspective, the length of titles and abstract is therefore of interest.

In our results, in 19 of the 107 areas considered there are differences between the median length of the abstract corresponding to the HCA and NHCA groups at a significance level of 1%. If a significance level of 5% is considered, the number of research areas showing differences between medians rises to 39. When large samples are considered, the equality-of-medians hypothesis is rejected at the 0.05 level in 57% of the research areas with more than 50 observations in each group. In most cases, the abstracts are longer in the HCA group.

However, in only 9 of the 107 areas are there differences between the median length of the title corresponding to the HCA and NHCA groups at a significance level of 1%. If a significance level of 5% is considered, the number of areas showing differences between medians rises to 19. When large samples are considered, the equality-of-medians hypothesis is rejected at the 0.05 level in 33% of the research areas with more than 50 observations in each group. In most cases, the length of the title is smaller in the HCA group.

These results provide empirical evidence in favour of the thesis that longer abstracts and shorter titles contribute to a paper achieving more citations.

*Number of pages and references*

In 56 of the 107 areas, there exist differences between the median number of pages corresponding to the HCA and NHCA groups at a significance level of 1%. If a significance level of 5% is considered, the number of research areas showing differences between medians rises to 62. When large samples are considered, the equality-of-medians hypothesis is rejected at the 0.01 level in 60% of research areas with more than 50 observations in each group (67% of areas at the 0.05 level). In most cases, the median number of pages is higher in the HCA group.

In 38 of the 107 areas, there are differences between the median number of references corresponding to the HCA and NHCA groups at a significance level of 1%. If a significance level of 5% is considered, the number of research areas showing differences between medians rises to 51. When large samples are considered, the equality-of-medians hypothesis is rejected at the 0.01 level in 67% of research areas with more than 50 observations (80% of areas at the 0.05 level). In all cases, the median number of references is higher in the HCA group.

These results are evidence that a higher number of pages and longer list of references contribute to a paper achieving more citations.

*Question form in title and abstract*

The title is very important for the visibility of a paper. However, the abstract is the key to persuading potential readers to finally read the paper. The frequency of the most common punctuation marks in the title and abstract are shown in Table 5. Question form titles appear in 2% of the HCA group and slightly less in the NHCA group (1.86%). However, question marks in the abstract are rare and only appear in approximately 1% of cases.

[Table 5 about here]

The colon in titles is quite frequent (23% in the HCA group and 18% in the NHCA group). This use is related to the size of the title because authors frequently use the colon to link sentences instead of other longer rhetorical figures. Therefore, the greater presence of this punctuation mark in the HCA group may explain the fact that titles are slightly shorter within this group.

A descriptive title maximizes the possibilities that readers correctly remember the arguments to rediscover what they are looking for. However, some authors adopt question form titles in the belief that they will be more attractive and increase the number of readers and citations. In an attempt to resolve this issue and determine whether question form titles influence the number of citations per year, a median test for this variable was performed, distinguishing between the HCA and NHCA groups (Table 6).

[Table 6 about here]

The results show differences in the HCA group in the median number of citations per year between papers titled in descriptive and question forms at a significance level of 1%. According to the results, in the HCA group the median number of citations per year is higher for papers with a descriptive title. No significant differences were found in the NHCA group.

*Citations per year in the highly cited article group*

It has been seen that there are significant differences in several characteristics between the two groups of articles, the HCA and the NHCA. It was also decided to determine whether there existed significant differences within the HCA group. For the purpose of brevity, we will limit ourselves to describing some of the variables graphically.

Scatterplots between citations per year and the variables that most affect the impact are shown in Figure 2. As can be observed, in the case of JIF percentile, the vast majority of the highly cited articles are published in journals of the first quartile (percentile above 75%). Furthermore, an exponential relation between the two variables can be clearly seen. Within the select group of articles in the first quartile, there are numerous cases with more than a hundred citations per year in the analysed period. In five cases the number of citations per year is more than five hundred. However, in the group of articles published in journals of the second quartile (percentile 50 to 75) only a small number of papers with more than a hundred citations are observed.

[Figure 2 about here]

For number of authors there are two different trends depending on a specific threshold. Up to about 10 authors, there is a positive effect on citations. However, above that level there is no clear effect on citations. Surprisingly, there are about twenty papers with more than a hundred authors and one of them close to one thousand authors.

Something similar happens with number of pages. There are two different trends. Up to around 15 pages, there is a positive effect on citations. However, above that level there is again no clear effect on citations. Perhaps surprisingly, there are many cases of articles more than fifty pages long.

Finally, as can be deduced from Figure 2, all of the ten most cited papers (with more than four hundred citations per year) were published in journals above the percentile 90, and only one had more than a hundred authors. In addition, three are over twenty-five pages long. The most cited document is a major collaboration of around five hundred authors. The second and third most cited articles are two very extensive documents of more than two hundred and fifty pages.

**Conclusions**

There is no strong evidence in the literature in favour of the thesis that some formal aspects contribute to achieving more citations for a paper. The results mainly depend on the methodology employed and there is no consensus in some aspects about the real effect on citations.

This large-scale study, both in terms of sample size and the number of areas considered, analyses metadata associated with the publications and concludes that some of them have a significant influence in explaining the impact of documents.

Although the number of authors might be expected to have an impact on both article length and number of citations, no correlation was found in this study between number of authors and number of pages, nor between number of authors and JIF.

Highly cited articles differ from the other articles in most bibliometric aspects. There are significant differences, below the 0.01 level, between highly cited and non-highly cited articles in many variables and areas. Highly cited articles are published in journals with a higher impact factor (33 percentile points above) and have more authors (25% more). Highly cited articles are also longer in terms of number of pages (10% higher) and bibliographical references (35% more). Finally, highly cited articles have slightly shorter titles (3% lower) but, contrastingly, longer abstracts (10% higher).

The practical implications of these results are related mainly to the impact of the publication journal and the impact of the article. The publication journal is very important in relation to the impact of the research because the journal impact factor percentile is usually a good measure of visibility and readership. From the point of view of research impact, it is preferable for titles to be descriptive and short, and the abstract to be the part that extensively describes the conclusions and methodological aspects.

**References**

Aksnes, D. W. (2003). Characteristics of highly cited papers. *Research Evaluation*, 12(3), 159–170.

Barnett, G. A., & Fink, E. L. (2008). Impact of the internet and scholar age distribution on academic citation age. *Journal of the American Society for Information Science and Technology*, 59(4), 526–534.

Bhandari, M., Busse, J., Devereaux, P. J., Montori, V. M., Swiontkowski, M., Tornetta Iii, P., & Schemitsch, E. H. (2007). Factors associated with citation rates in the orthopedic literature. *Canadian Journal of Surgery*, 50(2), 119–123.

Biscaro, C., & Giupponi, C. (2014). Co-authorship and bibliographic coupling network effects on citations. *PLoS One*, 9(6), e99502.

Bornmann, L., Schier, H., Marx, W., & Daniel, H. D. (2012). What factors determine citation counts of publications in chemistry besides their quality? *Journal of Informetrics*, 6(1), 11–18.

Buela-Casal, G., & Zych, I. (2010). Analysis of the relationship between the number of citations and the quality evaluated by experts in psychology journals. *Psicothema*, 22(2), 270–276.

Buter, R. K., & van Raan, A. F. J. (2011). Non-alphanumeric characters in titles of scientific publications: An analysis of their occurrence and correlation with citation impact. *Journal of Informetrics*, 5(4), 608–617.

Callaham, M., Wears, R. L., & Weber, E. (2002). Journal prestige, publication bias, and other characteristics associated with citation of published studies in peer-reviewed journals. *Journal of the American Medical Association*, 287(21), 2847–2850.

Chen, C. M. (2012). Predictive effects of structural variation on citation counts. *Journal of the American Society for Information Science and Technology*, 63(3), 431–449.

Costas, R., Bordons, M., Van Leeuwen, T. N., & Van Raan, A. F. J. (2009). Scaling rules in the science system: Influence of field-specific citation characteristics on the impact of individual researchers. *Journal of the American Society for Information Science and Technology*, 60(4), 740–753.

Dorta-González, P., & Dorta-González, M. I. (2013a). Comparing journals from different fields of science and social science through a JCR subject categories normalized impact factor. *Scientometrics*, 95(2), 645–672.

Dorta-González, P., & Dorta-González, M. I. (2013b). Impact maturity times and citation time windows: The 2-year maximum journal impact factor. *Journal of Informetrics,* 7(3), 593–602.

Dorta-González, P., Dorta-González, M. I., Santos-Peñate, D. R., & Suárez-Vega, R. (2014). Journal topic citation potential and between-field comparisons: The topic normalized impact factor. *Journal of Informetrics*, 8(2), 406–418.

Dorta-González, P., González-Betancor, S. M., & Dorta-González, M. I. (2017). Reconsidering the gold open access citation advantage postulate in a multidisciplinary context: an analysis of the subject categories in the Web of Science database 2009-2014. *Scientometrics*, 112(2), 877–901.

Dorta-González, P., & Santana-Jiménez, Y. (2018). Prevalence and citation advantage of gold open access in the subject areas of the Scopus database. *Research Evaluation*, 27(1), 1–15.

Falagas, M. E., Zarkali, A., Karageorgopoulos, D. E., Bardakas, V., & Mavros, M. N. (2013). The impact of article length on the number of future citations: A bibliometric analysis of general medicine journals. *PLoS One*, 8(2), e49476.

Fu, L. D., & Aliferis, C. F. (2010). Using content-based and bibliometric features for machine learning models to predict citation counts in the biomedical literature. *Scientometrics*, 85(1), 257–270.

Garfield, E. (2006). The history and meaning of the journal impact factor. *Journal of the American Medical Association*, 295, 90–93.

Garner, J., Porter, A. L., & Newman, N. C. (2014). Distance and velocity measures: Using citations to determine breadth and speed of research impact. *Scientometrics*, 100(3), 687–703.

Harwood, N. (2008). Publication outlets and their effect on academic writers' citations. *Scientometrics*, 77(2), 253–265.

Jacques, T. S., & Sebire, N. J. (2010). The impact of article titles on citation hits: An analysis of general and specialist medical journals. *JRSM Short Reports*, 1(1), 2.

Jamali, H. R., & Nikzad, M. (2011). Article title type and its relation with the number of downloads and citations. *Scientometrics*, 88(2), 653–661.

Patterson, M. S., & Harris, S. (2009). The relationship between reviewers' quality-scores and number of citations for papers published in the journal physics in medicine and biology from 2003-2005. *Scientometrics*, 80(2), 343–349.

Peng, T. Q., & Zhu, J. J. H. (2012). Where you publish matters most: A multilevel analysis of factors affecting citations of internet studies. *Journal of the American Society for Information Science and Technology*, 63(9), 1789–1803.

Rostami, F., Mohammadpoorasl, A., & Hajizadeh, M. (2014). The effect of characteristics of title on citation rates of articles. *Scientometrics*, 98(3), 2007–2010.

Royle, P., Kandala, N. B., Barnard, K., & Waugh, N. (2013). Bibliometrics of systematic reviews: Analysis of citation rates and journal impact factors. *Systematic Reviews*, 2, 74.

Stremersch, S., Verniers, I., & Verhoef, P. C. (2007). The quest for citations: Drivers of article impact. *Journal of Marketing*, 71(3), 171–193.

Subotic, S., & Mukherjee, B. (2014). Short and amusing: The relationship between title characteristics, downloads, and citations in psychology articles. *Journal of Information Science*, 40(1), 115–124.

Tahamtan, I., Askar Safipour, A., & Ahamdzadeh, K. (2016). Factors affecting number of citations: a comprehensive review of the literature. *Scientometrics*, 107(3), 1195–1225.

van Wesel, M., Wyatt, S., & ten Haaf, J. (2014). What a difference a colon makes: How superficial factors influence subsequent citation. *Scientometrics*, 98(3), 1601–1615.

Table 1: Pearson correlation coefficient by groups of articles between the variables N References and N Pages for the 10 research areas of largest sample size

|  | Sample size | | Pearson correlation coefficient | |
|---|---|---|---|---|
| Area | HCA | NHCA | HCA | NHCA |
| Biochemistry & Molecular Biology | 197 | 180 | 0.62 | 0.60 |
| Business & Economics | 171 | 101 | 0.39 | 0.22 |
| Chemistry | 719 | 492 | 0.73 | 0.41 |
| Computer Science | 149 | 186 | 0.58 | 0.39 |
| Engineering | 466 | 597 | 0.59 | 0.34 |
| Environmental Sciences & Ecology | 265 | 192 | 0.63 | 0.53 |
| General & Internal Medicine | 285 | 116 | 0.74 | 0.72 |
| Materials Science | 453 | 288 | 0.67 | 0.32 |
| Mathematics | 280 | 269 | 0.42 | 0.36 |
| Physics | 550 | 636 | 0.34 | 0.51 |

Table 2: Medians by groups of papers, and equality-of-medians tests in 6 bibliometric aspects for aggregated research areas

|  | Highly Cited Articles (n=4,956) | Non-Highly Cited Articles (n=4,998) | Non-parametric Test |
|---|---|---|---|
|  | Median | Median | p-value |
| **N Authors** | 5 | 4 | .00 |
| **N Title characters** | 90 | 93 | .00 |
| **N Abstract characters** | 1,278 | 1,160 | .00 |
| **N References** | 41 | 30 | .00 |
| **N Pages** | 10 | 9 | .00 |
| **JIF Percentile** | 91 | 58 | .00 |

[Table 3 is at the end of the document]

Table 4: Number of areas, from all 107 research areas and the 30 areas with more than 50 papers in both groups, where there are significant differences between groups of articles

| Areas | p-value | | N Authors | N Title characters | N Abstract characters | N References | N Pages | JIF Percentile |
|---|---|---|---|---|---|---|---|---|
| 107 | p ≤ .01 | HCA > NHCA | 36 | 1 | 15 | 38 | 43 | 79 |
| | | HCA < NHCA | 0 | 8 | 4 | 0 | 8 | 0 |
| | | Total (of 107) | **36** | **9** | **19** | **38** | **56** | **79** |
| | p ≤ .05 | HCA > NHCA | 51 | 6 | 30 | 51 | 49 | 86 |
| | | HCA < NHCA | 0 | 13 | 9 | 0 | 8 | 0 |
| | | Total (of 107) | **51** | **19** | **39** | **51** | **62** | **86** |
| 30 | p ≤ .01 | HCA > NHCA | 21 | 0 | 5 | 20 | 14 | 30 |
| | | HCA < NHCA | 0 | 5 | 3 | 0 | 4 | 0 |
| | | Total (of 30) | **21** | **5** | **8** | **20** | **18** | **30** |
| | p ≤ .05 | HCA > NHCA | 24 | 2 | 10 | 24 | 15 | 30 |
| | | HCA < NHCA | 0 | 8 | 7 | 0 | 4 | 0 |
| | | Total (of 30) | **24** | **10** | **17** | **24** | **20** | **30** |

Table 5: Frequency of most common punctuation marks in title and abstract

| | | Highly Cited Articles (n=4,956) | | Non-Highly Cited Articles (n=4,998) | |
|---|---|---|---|---|---|
| | Character | Frequency | Percentage | Frequency | Percentage |
| **Title** | ? | 99 | 2.00% | 93 | 1.86% |
| | : | 1,153 | 23.26% | 916 | 18.33% |
| | . | 24 | 0.48% | 48 | 0.96% |
| **Abstract** | ? | 48 | 0.97% | 46 | 0.92% |

Table 6: Median of citations per year and median test by groups for titles in question and descriptive forms

| | | Question form | Descriptive | Median Test p-value |
|---|---|---|---|---|
| **HCA** | **Cites per year** | 19 | 25.5 | 0.00 |
| | **Frequency** | 99 | 4856 | |
| **NHCA** | **Cites per year** | 0.75 | 1 | 0.17 |
| | **Frequency** | 93 | 4905 | |

Figure 1: Medians by groups of articles in 107 research areas

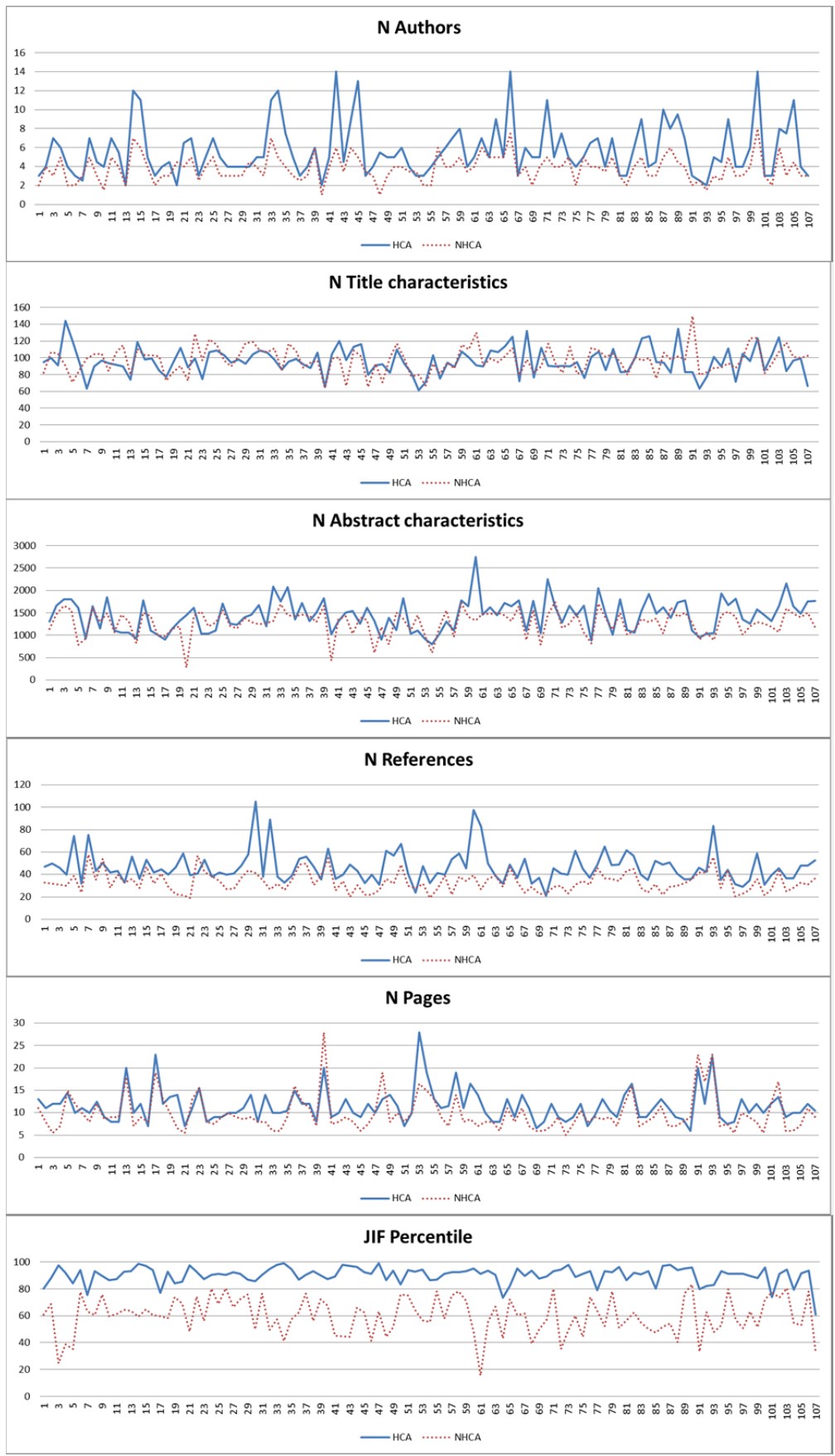

Figure 2: Scatterplots for the highly cited articles group

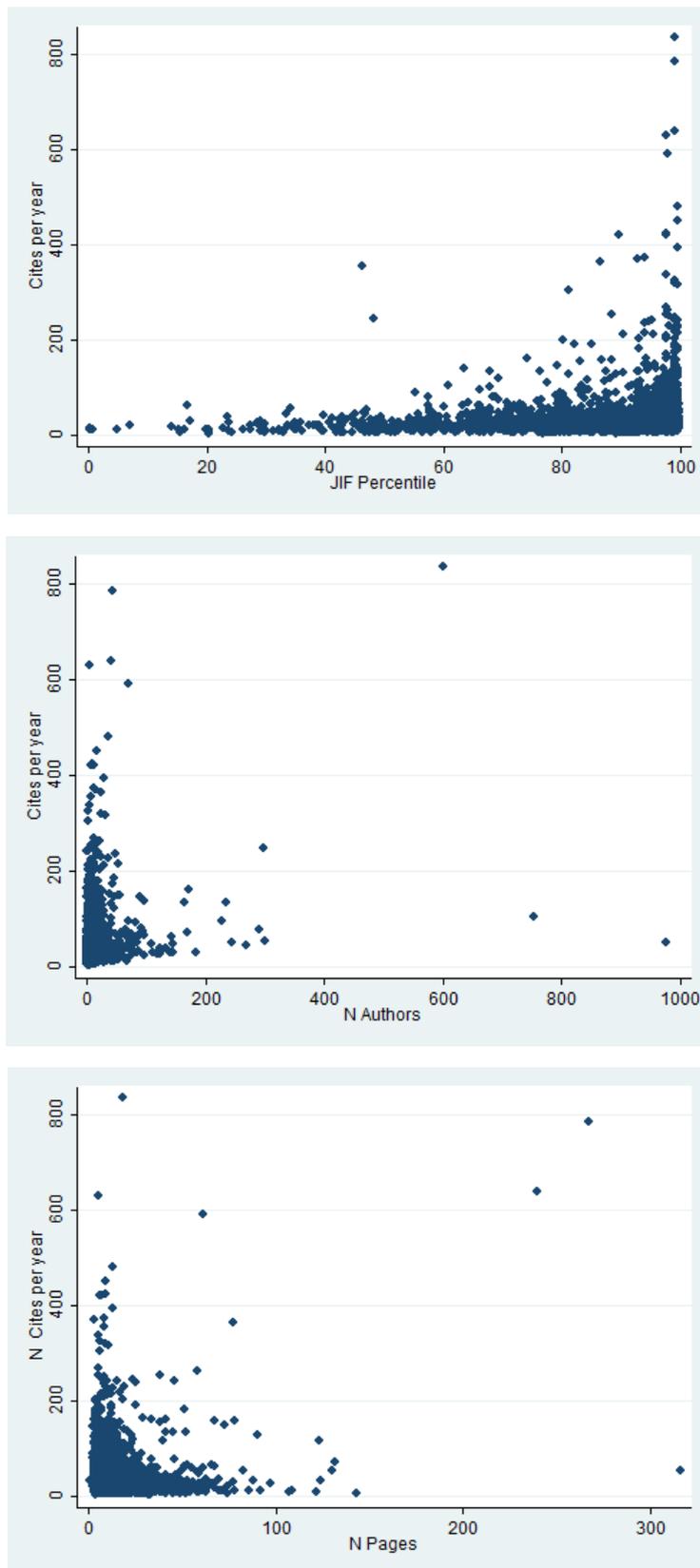

Table 3: Medians by groups and equality-of-medians tests (green colour below 0.05 level) for 6 bibliometric aspects in 107 research areas (Source: Web of Science)

| | | n | | N Authors | | | N Title characters | | | N Abstract characters | | | N References | | | N Pages | | | JIF Percentile | | |
|---|---|---|---|---|---|---|---|---|---|---|---|---|---|---|---|---|---|---|---|---|---|
| | Area | HCA | NHCA | HCA | NHCA | p-value | HCA | NHCA | p-value | HCA | NHCA | p-value | HCA | NHCA | p-value | HCA | NHCA | p-value | HCA | NHCA | p-value |
| 1 | Acoustics | 5 | 15 | 3 | 2 | 0.176 | 95 | 82 | 0.121 | 1310 | 1137 | 0.606 | 47 | 33 | 0.606 | 13 | 11 | 0.121 | 80 | 60.94 | 0.37 |
| 2 | Agriculture | 56 | 127 | 4 | 4 | 0.751 | 100 | 107 | 0.145 | 1655.5 | 1467 | 0.009 | 50 | 32 | 0 | 11 | 8 | 0 | 87.89 | 68.75 | 0 |
| 3 | Allergy | 5 | 4 | 7 | 3 | 0.294 | 91 | 105 | 0.764 | 1801 | 1664.5 | 0.294 | 46 | 31 | 0.294 | 12 | 5.5 | 0.058 | 97.5 | 24.73 | 0 |
| 4 | Anesthesiology | 9 | 12 | 6 | 5 | 0.528 | 144 | 90.5 | 0.13 | 1801 | 1570.5 | 0.528 | 40 | 30 | 0.017 | 12 | 7 | 0.001 | 91.94 | 38.86 | 0.02 |
| 5 | Anthropology | 8 | 9 | 4 | 2 | 0.03 | 121 | 71 | 0.229 | 1613.5 | 790 | 0.03 | 74.5 | 39 | 0.229 | 14.5 | 15 | 0.402 | 84.11 | 35.12 | 0 |
| 6 | Automation & Control Systems | 63 | 19 | 3 | 2 | 0.07 | 93 | 86 | 0.432 | 924 | 932 | 0.702 | 32 | 24 | 0.286 | 10 | 12 | 0.202 | 93.8 | 77.78 | 0 |
| 7 | Behavioral Sciences | 8 | 7 | 2.5 | 3 | 0.833 | 63 | 99 | 0.005 | 1645.5 | 1610 | 0.782 | 75.5 | 58 | 0.189 | 11 | 10 | 0.782 | 75.45 | 63.24 | 0.4 |
| 8 | Biochemistry & Molecular Biology | 197 | 180 | 7 | 5 | 0.001 | 90 | 104.5 | 0 | 1146 | 1293 | 0.012 | 43 | 35 | 0.006 | 10 | 8 | 0.059 | 93.28 | 60.6 | 0 |
| 9 | Biodiversity & Conservation | 12 | 9 | 4.5 | 3 | 0.445 | 96.5 | 105 | 0.056 | 1844.5 | 1499 | 0.044 | 50.5 | 54 | 0.528 | 12.5 | 12 | 0.195 | 89.6 | 75.88 | 0.03 |
| 10 | Biomedical Social Sciences | 12 | 4 | 4 | 1.5 | 0.042 | 93.5 | 84 | 1 | 1101.5 | 1065.5 | 1 | 42 | 28 | 1 | 9 | 8.5 | 0.712 | 86.31 | 59.85 | 0.43 |
| 11 | Biophysics | 14 | 40 | 7 | 5 | 0.214 | 91.5 | 106 | 0.062 | 1059.5 | 1455 | 0.013 | 43 | 40 | 0.872 | 8 | 9 | 0.347 | 87.26 | 61.28 | 0.01 |
| 12 | Biotechnology & Applied Microbiology | 66 | 89 | 5.5 | 4 | 0.108 | 90 | 115 | 0 | 1057 | 1307 | 0.004 | 33 | 33 | 0.983 | 8 | 9 | 0.219 | 92.7 | 64.53 | 0 |
| 13 | Business & Economics | 171 | 101 | 2 | 2 | 0.194 | 74 | 80 | 0.211 | 925 | 816 | 0.259 | 56 | 36 | 0 | 20 | 18 | 0.109 | 93.01 | 63.54 | 0 |
| 14 | Cardiovascular System & Cardiology | 146 | 78 | 12 | 7 | 0 | 119 | 110.5 | 0.04 | 1776.5 | 1508 | 0.002 | 36 | 28 | 0.007 | 10 | 7 | 0 | 98.49 | 59.11 | 0 |
| 15 | Cell Biology | 94 | 67 | 11 | 6 | 0 | 98 | 103 | 0.821 | 1101 | 1424 | 0 | 53 | 47 | 0.053 | 12 | 9 | 0 | 97.26 | 64.84 | 0 |
| 16 | Chemistry | 719 | 492 | 5 | 4 | 0 | 99 | 103 | 0.278 | 1004 | 994.5 | 0.498 | 42 | 32 | 0 | 7 | 8 | 0.005 | 94.03 | 60.47 | 0 |
| 17 | Communication | 14 | 7 | 3 | 2 | 0.186 | 85 | 102 | 0.122 | 904.5 | 979 | 0.537 | 44.5 | 41 | 0.35 | 23 | 19 | 0.031 | 76.94 | 59.54 | 0.06 |
| 18 | Computer Science | 149 | 186 | 4 | 3 | 0 | 77 | 72.5 | 0.427 | 1141 | 1141 | 0.971 | 40 | 29 | 0 | 12 | 13 | 0.152 | 92.7 | 58.08 | 0 |
| 19 | Construction & Building Technology | 8 | 30 | 4.5 | 3 | 0.078 | 95.5 | 83.5 | 0.335 | 1322 | 1199.5 | 0.111 | 46.5 | 22.5 | 0.053 | 13.5 | 10 | 0.335 | 83.99 | 74 | 0.08 |
| 20 | Cristallography | 5 | 8 | 2 | 4.5 | 0.279 | 112 | 90.5 | 0.429 | 1446 | 270.5 | 0.429 | 59 | 21.5 | 0.053 | 14 | 6.5 | 0.053 | 85.09 | 69.3 | 0.21 |
| 21 | Dermatology | 12 | 24 | 6.5 | 4 | 0.024 | 88.5 | 72.5 | 0.034 | 1608.5 | 1441 | 0.48 | 39.5 | 19 | 0.002 | 7 | 5.5 | 0.236 | 97.54 | 48.38 | 0 |
| 22 | Developmental Biology | 5 | 9 | 7 | 5 | 0.577 | 99 | 129 | 0.198 | 1032 | 1526 | 0.005 | 41 | 57 | 0.577 | 11 | 13 | 0.577 | 92.81 | 74.39 | 0.33 |
| 23 | Education & Educational Research | 30 | 38 | 3 | 2.5 | 0.218 | 74.5 | 97 | 0.015 | 1032.5 | 1203 | 0.329 | 53 | 42.5 | 0.329 | 15.5 | 15.5 | 1 | 87.15 | 56.3 | 0 |
| 24 | Electrochemistry | 65 | 60 | 5 | 4 | 0.001 | 107 | 122 | 0.129 | 1102 | 1280.5 | 0.025 | 38 | 38.5 | 0.932 | 8 | 8 | 0.273 | 90.4 | 80.09 | 0 |
| 25 | Endocrinology & Metabolism | 58 | 70 | 7 | 5 | 0 | 108.5 | 115.5 | 0.287 | 1712 | 1566 | 0.076 | 42 | 34 | 0.033 | 9 | 7.5 | 0.039 | 91.22 | 68.78 | 0 |
| 26 | Energy & Fuels | 218 | 108 | 5 | 3 | 0 | 103 | 96.5 | 0.272 | 1252 | 1207.5 | 0.347 | 40 | 27 | 0 | 9 | 9 | 0.861 | 90.4 | 80.36 | 0.25 |
| 27 | Engineering | 466 | 597 | 4 | 3 | 0 | 94 | 90 | 0.09 | 1233.5 | 1156 | 0.045 | 41 | 27 | 0 | 10 | 10 | 0.508 | 92.53 | 66.29 | 0 |
| 28 | Environmental Sciences & Ecology | 265 | 192 | 4 | 3 | 0.001 | 98 | 101.5 | 0.239 | 1397 | 1386 | 0.597 | 48 | 37 | 0 | 10 | 9 | 0.472 | 91.21 | 72.16 | 0 |
| 29 | Evolutionary Biology | 33 | 10 | 4 | 3 | 0.481 | 93 | 117.5 | 0.42 | 1457 | 1291 | 0.174 | 58 | 43.5 | 0.174 | 11 | 8.5 | 0.385 | 87.01 | 75.85 | 0.71 |
| 30 | Fisheries | 5 | 12 | 4 | 4.5 | 0.707 | 104 | 118.5 | 0.149 | 1676 | 1249 | 0.079 | 105 | 41.5 | 0.079 | 14 | 9 | 0.013 | 85.58 | 49.66 | 0.01 |
| 31 | Food Science & Technology | 73 | 82 | 5 | 4 | 0.023 | 109 | 108.5 | 0.932 | 1202 | 1271.5 | 0.466 | 38 | 34.5 | 0.023 | 8 | 8 | 0.5 | 90.29 | 76.53 | 0 |
| 32 | Forestry | 13 | 31 | 5 | 3 | 0.027 | 106 | 107 | 0.515 | 2082 | 1314 | 0.008 | 89 | 26.5 | 0 | 14 | 8 | 0 | 94.7 | 49.43 | 0 |
| 33 | Gastroenterology & Hepatology | 55 | 55 | 11 | 7 | 0 | 98 | 111 | 0.182 | 1759 | 1690 | 0.182 | 38 | 32 | 0.086 | 10 | 6 | 0 | 98.08 | 57.17 | 0 |
| 34 | General & Internal Medicine | 285 | 116 | 12 | 5 | 0 | 86 | 85 | 0.928 | 2079 | 1450.5 | 0 | 33 | 26 | 0 | 10 | 6 | 0 | 99.01 | 41.06 | 0 |

| # | Field | | | | | | | | | | | | | | | | | | | |
|---|---|---|---|---|---|---|---|---|---|---|---|---|---|---|---|---|---|---|---|---|
| 35 | Genetics & Heredity | 46 | 47 | 7.5 | 4 | 0.049 | 95.5 | 117 | 0.003 | 1356 | 1418 | 0.467 | 39 | 35 | 0.605 | 10.5 | 9 | 0 | 94.66 | 57.35 | 0 |
| 36 | Geochemistry & Geophysics | 20 | 31 | 5 | 3 | 0.018 | 98.5 | 108 | 0.645 | 1714 | 1455 | 0.067 | 54 | 49 | 0.361 | 15 | 16 | 0 | 86.74 | 62.35 | 0 |
| 37 | Geography | 27 | 4 | 3 | 2.5 | 0.17 | 92 | 88.5 | 0.316 | 1322 | 1431.5 | 0.945 | 56 | 50 | 0.945 | 12 | 12.5 | 0 | 90.26 | 76.49 | 0.39 |
| 38 | Geology | 71 | 78 | 4 | 3 | 0.02 | 88 | 94 | 0.277 | 1507 | 1293.5 | 0.027 | 47 | 30.5 | 0.012 | 12 | 11 | 0 | 93.21 | 56.29 | 0 |
| 39 | Geriatrics & Gerontology | 16 | 18 | 6 | 6 | 0.515 | 106 | 97 | 0.492 | 1826 | 1655 | 0.169 | 35.5 | 39.5 | 0.492 | 8 | 7 | 0 | 90 | 72.17 | 0.04 |
| 40 | Government & Law | 22 | 34 | 2 | 1 | 0.003 | 65 | 64 | 0.83 | 1025.5 | 429.5 | 0.029 | 63 | 54.5 | 0.584 | 20 | 28 | 0 | 87.41 | 67.69 | 0 |
| 41 | Health Care Sciences & Services | 53 | 24 | 5 | 4 | 0.429 | 104 | 99 | 0.678 | 1302 | 1349.5 | 0.939 | 36 | 25.5 | 0.037 | 9 | 7.5 | 0 | 89.2 | 44.96 | 0 |
| 42 | Hematology | 53 | 24 | 14 | 6 | 0 | 120 | 100.5 | 0.364 | 1507 | 1424 | 0.364 | 40 | 34 | 0.364 | 10 | 8 | 0 | 97.86 | 44.49 | 0 |
| 43 | Imaging Science & Photographic Technology | 12 | 12 | 4.5 | 3.5 | 0.083 | 97.5 | 66.5 | 0.102 | 1544.5 | 1038.5 | 0.102 | 49 | 20 | 0.001 | 13 | 9 | 0.041 | 97.3 | 44.21 | 0 |
| 44 | Immunology | 49 | 58 | 9 | 6 | 0.016 | 113 | 108.5 | 0.777 | 1261 | 1394.5 | 0.378 | 43 | 30.5 | 0.066 | 10 | 8 | 0 | 96.42 | 65.85 | 0 |
| 45 | Infectious Diseases | 30 | 61 | 13 | 5 | 0.009 | 116 | 102 | 0.207 | 1608 | 1343 | 0.334 | 32.5 | 22 | 0.004 | 9 | 6 | 0 | 92.28 | 62.54 | 0 |
| 46 | Information Science & Library Science | 21 | 8 | 3 | 3.5 | 0.73 | 80 | 65 | 0.624 | 1316 | 611.5 | 0.017 | 40 | 22 | 0.122 | 12 | 7.5 | 0 | 91.15 | 41.06 | 0 |
| 47 | Instruments & Instrumentation | 27 | 49 | 4 | 3 | 0.731 | 91 | 93 | 0.705 | 908 | 1187 | 0.008 | 31 | 26 | 0.093 | 10 | 10 | 0 | 98.96 | 63.13 | 0 |
| 48 | International Relations | 10 | 9 | 5.5 | 1 | 0.096 | 92 | 71 | 0.245 | 1382.5 | 791 | 0.037 | 61 | 36 | 0.037 | 13 | 19 | 0 | 86.57 | 44.19 | 0 |
| 49 | Life Sciences & Biomedicine - Other Topics | 35 | 36 | 5 | 3 | 0.001 | 82 | 100.5 | 0.122 | 1118 | 1501.5 | 0.122 | 57 | 32 | 0 | 14 | 8 | 0 | 93.61 | 51.26 | 0 |
| 50 | Marine & Freshwater Biology | 14 | 27 | 5 | 4 | 0.318 | 110 | 117 | 0.585 | 1830 | 1365 | 0.153 | 67.5 | 49 | 0.153 | 11.5 | 10 | 0 | 83.17 | 75.93 | 0.19 |
| 51 | Materials Science | 453 | 288 | 6 | 4 | 0 | 93 | 99.5 | 0.021 | 1036 | 1102 | 0.022 | 41 | 30 | 0 | 7 | 8 | 0 | 94.03 | 75.23 | 0 |
| 52 | Mathematical & Computational Biology | 48 | 18 | 4 | 3.5 | 0.641 | 82 | 77.5 | 0.58 | 1106.5 | 1449 | 0.097 | 24 | 27.5 | 0.58 | 10 | 10 | 0 | 92.7 | 64.53 | 0 |
| 53 | Mathematical Methods In Social Sciences | 20 | 8 | 3 | 3.5 | 0.112 | 61.5 | 79.5 | 0.403 | 910.5 | 983.5 | 1 | 47.5 | 32 | 0.023 | 28 | 16.5 | 0 | 94.35 | 56.53 | 0.01 |
| 54 | Mathematics | 280 | 269 | 3 | 2 | 0 | 71 | 67 | 0.112 | 793 | 613 | 0.002 | 32.5 | 19 | 0 | 19 | 15 | 0 | 86.42 | 55.4 | 0 |
| 55 | Mechanics | 64 | 83 | 4 | 2 | 0 | 103 | 93 | 0.224 | 1038 | 1192 | 0.054 | 41.5 | 27 | 0.004 | 13 | 13 | 0 | 86.81 | 78.15 | 0 |
| 56 | Medical Informatics | 15 | 5 | 5 | 6 | 0.292 | 75 | 81 | 0.606 | 1302 | 1551 | 0.121 | 40 | 39 | 0.606 | 11 | 9 | 0 | 91.15 | 57.5 | 0.06 |
| 57 | Metallurgy & Metallurgical Engineering | 6 | 35 | 6 | 4 | 0.175 | 94 | 94 | 0.948 | 1101.5 | 967 | 0.067 | 53.5 | 22 | 0.224 | 11.5 | 7 | 0 | 92.24 | 75.23 | 0.08 |
| 58 | Meteorology & Atmospheric Sciences | 57 | 38 | 7 | 4 | 0 | 89 | 87.5 | 0.867 | 1782 | 1721 | 0.738 | 59 | 38 | 0.003 | 19 | 14 | 0 | 92.26 | 78.23 | 0 |
| 59 | Microbiology | 44 | 70 | 8 | 5 | 0.004 | 107.5 | 115.5 | 0.442 | 1647 | 1415 | 0.021 | 45.5 | 34 | 0.002 | 11 | 8 | 0 | 93.05 | 71.14 | 0 |
| 60 | Mineralogy | 8 | 4 | 4 | 3.5 | 0.665 | 100.5 | 110 | 1 | 2751.5 | 1333.5 | 0.014 | 97.5 | 39.5 | 0.014 | 16.5 | 8.5 | 0.408 | 94.95 | 48.12 | 0.01 |
| 61 | Mycology | 11 | 4 | 5 | 4 | 0.31 | 91 | 130 | 0.185 | 1459 | 1487.5 | 0.876 | 83 | 26.5 | 0.029 | 14 | 7 | 0 | 91.38 | 15.52 | 0.05 |
| 62 | Neurosciences & Neurology | 140 | 127 | 7 | 6 | 0.027 | 90 | 91 | 0.67 | 1624 | 1465 | 0.197 | 50 | 36 | 0.003 | 10 | 8 | 0 | 93.65 | 55.15 | 0 |
| 63 | Nutrition & Dietetics | 70 | 23 | 5 | 5 | 0.768 | 108.5 | 99 | 0.856 | 1447.5 | 1504 | 0.435 | 39 | 39 | 0.951 | 8 | 8 | 0 | 90.29 | 66.67 | 0.01 |
| 64 | Obstetrics & Gynecology | 11 | 29 | 9 | 5 | 0.208 | 107 | 95 | 0.288 | 1722 | 1444 | 0.077 | 32 | 29 | 0.583 | 8 | 6 | 0 | 73.27 | 43.13 | 0 |
| 65 | Oceanography | 13 | 20 | 5 | 5 | 0.727 | 114 | 102 | 0.226 | 1646 | 1312 | 0.226 | 49 | 46.5 | 0.619 | 13 | 11 | 0 | 82.25 | 72.16 | 0.13 |
| 66 | Oncology | 178 | 58 | 14 | 7.5 | 0 | 125 | 112 | 0.041 | 1779.5 | 1644.5 | 0.034 | 37 | 33.5 | 0.081 | 9 | 8 | 0 | 95.07 | 60.33 | 0 |
| 67 | Operations Research & Management Science | 19 | 41 | 3 | 3 | 0.873 | 72 | 79 | 0.781 | 1095 | 897 | 0.405 | 54 | 24 | 0 | 14 | 11 | 0 | 89.56 | 61.59 | 0.01 |
| 68 | Ophthalmology | 7 | 39 | 6 | 4 | 0.355 | 132 | 98 | 0.681 | 1771 | 1562 | 0.175 | 32 | 29 | 0.592 | 11 | 7 | 0 | 93.75 | 39.33 | 0.01 |
| 69 | Optics | 54 | 71 | 5 | 2 | 0 | 76.5 | 83 | 0.488 | 1040.5 | 789 | 0 | 37 | 23 | 0.016 | 6.5 | 6 | 0 | 87.83 | 49.88 | 0 |
| 70 | Orthopedics | 9 | 52 | 5 | 4 | 0.007 | 112 | 89.5 | 0.063 | 2248 | 1446 | 0.063 | 21 | 22.5 | 0.758 | 8 | 6 | 0 | 89.12 | 56.87 | 0.01 |
| 71 | Parasitology | 14 | 29 | 11 | 5 | 0 | 90.5 | 117 | 0.232 | 1641.5 | 1738 | 0.586 | 45.5 | 29 | 0.007 | 12 | 7 | 0 | 93.05 | 80.06 | 0.01 |
| 72 | Pathology | 9 | 22 | 5 | 4 | 0.193 | 90 | 93 | 0.397 | 1277 | 1159 | 0.193 | 41 | 30 | 0.193 | 9 | 9 | 0 | 94.23 | 35.22 | 0 |

| # | Field | | | | | | | | | | | | | | | | | | | |
|---|---|---|---|---|---|---|---|---|---|---|---|---|---|---|---|---|---|---|---|---|
| 73 | Pediatrics | 16 | 64 | 7.5 | 4 | 0.057 | 90.5 | 82 | 0.094 | 1655.5 | 1253.5 | 0.001 | 40 | 23 | 0.004 | 8 | 5 | 0 | 97.92 | 49.47 | 0 |
| 74 | Pharmacology & Pharmacy | 57 | 122 | 5 | 5 | 0.269 | 90 | 113 | 0.001 | 1438 | 1477 | 0.667 | 61 | 31 | 0 | 9 | 7.5 | 0.001 | 88.74 | 60.16 | 0 |
| 75 | Physical Geography | 15 | 14 | 4 | 2 | 0.035 | 95 | 81 | 0.837 | 1662 | 1057.5 | 0.005 | 45 | 34 | 0.573 | 12 | 10.5 | 0.191 | 91.21 | 44.21 | 0.01 |
| 76 | Physics | 550 | 636 | 5 | 5 | 0.093 | 76 | 85 | 0 | 879 | 824 | 0.092 | 37 | 31 | 0 | 7 | 8 | 0 | 93.04 | 73.9 | 0 |
| 77 | Physiology | 6 | 37 | 6.5 | 4 | 0.068 | 101 | 112 | 0.413 | 2049 | 1710 | 0.007 | 48.5 | 46 | 0.286 | 9.5 | 9 | 0.853 | 78.84 | 63.69 | 0.14 |
| 78 | Plant Sciences | 88 | 90 | 7 | 4 | 0 | 107.5 | 109 | 0.764 | 1458 | 1381.5 | 0.025 | 65 | 36.5 | 0 | 13 | 8.5 | 0 | 93.06 | 52.19 | 0 |
| 79 | Polymer Science | 14 | 68 | 4 | 3.5 | 0.45 | 85.5 | 101 | 0.24 | 1007 | 1123.5 | 0.557 | 48.5 | 36 | 0.063 | 10.5 | 9 | 0.373 | 92.35 | 78.18 | 0.04 |
| 80 | Psychiatry | 64 | 41 | 7 | 5 | 0.002 | 110.5 | 105 | 0.243 | 1806 | 1499 | 0.003 | 49 | 34 | 0.034 | 9 | 7 | 0.016 | 96.34 | 50.8 | 0 |
| 81 | Psychology | 98 | 75 | 3 | 3 | 0.436 | 83 | 94 | 0.043 | 1115 | 1007 | 0.054 | 61.5 | 43 | 0.001 | 14 | 12 | 0.137 | 86.57 | 56.47 | 0 |
| 82 | Public Administration | 18 | 23 | 3 | 2 | 0.026 | 83.5 | 80 | 0.678 | 1061.5 | 1101 | 0.623 | 57 | 45 | 0.89 | 16.5 | 16 | 0.89 | 92.16 | 62.23 | 0 |
| 83 | Public, Environmental & Occupational Health | 151 | 93 | 6 | 4 | 0 | 100 | 101 | 0.62 | 1541 | 1370 | 0.048 | 40 | 28 | 0.002 | 9 | 7 | 0.007 | 90.94 | 54.99 | 0 |
| 84 | Radiology, Nuclear Medicine & Medical Imaging | 34 | 55 | 9 | 5 | 0.001 | 123.5 | 97 | 0.03 | 1918.5 | 1289 | 0 | 35 | 24 | 0.015 | 9 | 8 | 0.42 | 93.09 | 50.16 | 0 |
| 85 | Rehabilitation | 5 | 22 | 4 | 3 | 0.438 | 126 | 100 | 0.557 | 1477 | 1381.5 | 0.557 | 52 | 31.5 | 0.114 | 11 | 9 | 0.114 | 80.17 | 47.57 | 0.1 |
| 86 | Remote Sensing | 12 | 16 | 4.5 | 3 | 0.172 | 95 | 75.5 | 0.445 | 1630 | 1038.5 | 0.127 | 49 | 22 | 0.002 | 13 | 11.5 | 0.445 | 97.3 | 51.9 | 0 |
| 87 | Research & Experimental Medicine | 69 | 55 | 10 | 5 | 0 | 94 | 106 | 0.032 | 1391 | 1623 | 0.047 | 51 | 29 | 0 | 11 | 7 | 0 | 97.98 | 54.33 | 0 |
| 88 | Respiratory System | 33 | 23 | 8 | 6 | 0.017 | 82 | 98 | 0.415 | 1728 | 1415 | 0.014 | 41 | 30 | 0.001 | 9 | 7 | 0.074 | 93.97 | 40.44 | 0 |
| 89 | Rheumatology | 16 | 18 | 9.5 | 4.5 | 0.006 | 134.5 | 102 | 0.006 | 1773 | 1504 | 0.039 | 35.5 | 32.5 | 0.492 | 8.5 | 8 | 0.746 | 95.31 | 76.56 | 0 |
| 90 | Science & Technology - Other Topics | 819 | 190 | 7 | 4 | 0 | 83 | 98 | 0 | 1106 | 1276.5 | 0 | 36 | 35.5 | 0.62 | 6 | 9 | 0 | 96.03 | 83.33 | 0 |
| 91 | Social Issues | 5 | 5 | 3 | 2 | 0.058 | 83 | 150 | 0.527 | 956 | 907 | 0.527 | 46 | 42 | 0.527 | 20 | 23 | 0.527 | 79.59 | 33.4 | 0.06 |
| 92 | Social Sciences - Other Topics | 14 | 32 | 2.5 | 2.5 | 1 | 63 | 80 | 0.522 | 1037 | 1051.5 | 1 | 42.5 | 42 | 1 | 12 | 17 | 0.2 | 82.2 | 62.83 | 0.01 |
| 93 | Sociology | 6 | 16 | 2 | 1.5 | 0.24 | 78 | 83 | 0.338 | 1048.5 | 894.5 | 0.338 | 83.5 | 55.5 | 0.056 | 22.5 | 23 | 0.856 | 83.07 | 47.99 | 0 |
| 94 | Sport Sciences | 7 | 23 | 5 | 3 | 0.195 | 101 | 88 | 0.666 | 1927 | 1445 | 0.195 | 35 | 28 | 0.195 | 9 | 7 | 0.195 | 93.29 | 53.11 | 0.03 |
| 95 | Substance Abuse | 10 | 6 | 4.5 | 2.5 | 0.182 | 90 | 88.5 | 1 | 1666.5 | 1532.5 | 0.302 | 44 | 44 | 1 | 7.5 | 7.5 | 1 | 91.37 | 79.86 | 0.04 |
| 96 | Surgery | 46 | 126 | 9 | 5 | 0 | 111.5 | 93.5 | 0.031 | 1811 | 1409 | 0 | 31.5 | 20.5 | 0.001 | 8 | 5.5 | 0 | 91.09 | 56.87 | 0 |
| 97 | Tele26s | 36 | 49 | 4 | 3 | 0.125 | 71.5 | 88 | 0.018 | 1349.5 | 1006 | 0 | 29 | 23 | 0.274 | 13 | 10 | 0.006 | 91.25 | 50.61 | 0 |
| 98 | Thermodynamics | 48 | 52 | 4 | 3 | 0.029 | 105 | 100 | 0.321 | 1254.5 | 1199.5 | 0.109 | 34.5 | 26 | 0.005 | 10 | 9 | 0.86 | 89.61 | 63.36 | 0.11 |
| 99 | Toxicology | 15 | 37 | 6 | 4 | 0.202 | 96 | 123 | 0.126 | 1577 | 1298 | 0.126 | 59 | 36 | 0 | 12 | 8 | 0.001 | 88.2 | 51.29 | 0.01 |
| 100 | Transplantation | 7 | 10 | 14 | 8 | 0.008 | 123 | 124 | 0.772 | 1452 | 1258 | 0.092 | 31 | 21.5 | 0.008 | 10 | 5.5 | 0 | 96.12 | 72.2 | 0 |
| 101 | Transportation | 10 | 10 | 3 | 3 | 1 | 85 | 81.5 | 1 | 1319.5 | 1189.5 | 0.371 | 39.5 | 26.5 | 0.074 | 12 | 12 | 0.653 | 73.76 | 77.49 | 0.37 |
| 102 | Urban Studies | 8 | 13 | 3 | 2 | 0.248 | 103.5 | 93 | 0.864 | 1652 | 1073 | 0.284 | 45.5 | 43 | 0.864 | 13.5 | 17 | 0.195 | 91.21 | 73.32 | 0.07 |
| 103 | Urology & Nephrology | 28 | 45 | 8 | 6 | 0.016 | 124.5 | 107 | 0.566 | 2154.5 | 1596 | 0 | 36.5 | 25 | 0.001 | 9 | 6 | 0 | 94.16 | 80.39 | 0 |
| 104 | Veterinary Sciences | 12 | 41 | 7.5 | 3 | 0.007 | 84 | 118 | 0.058 | 1644.5 | 1480 | 0.465 | 36.5 | 28 | 0.302 | 10 | 6 | 0.012 | 79.2 | 54.56 | 0.04 |
| 105 | Virology | 21 | 30 | 11 | 4.5 | 0 | 97 | 101 | 0.461 | 1481 | 1404.5 | 0.332 | 48 | 33 | 0.019 | 10 | 7 | 0.304 | 91.44 | 53.03 | 0 |
| 106 | Water Resources | 29 | 55 | 4 | 3 | 0.054 | 99 | 100 | 0.818 | 1754 | 1509 | 0.012 | 48 | 31 | 0 | 12 | 11 | 0.386 | 93.73 | 78.07 | 0 |
| 107 | Zoology | 8 | 26 | 3 | 3 | 0.702 | 66.5 | 102.5 | 0.002 | 1769 | 1174 | 0.106 | 52.5 | 36.5 | 0.419 | 10.5 | 9 | 0.562 | 60.91 | 32.5 | 0 |